\documentclass[aps,prl,reprint,groupedaddress]{revtex4-2}
\usepackage{lineno}

\usepackage{graphicx}
\usepackage{amsmath}
\usepackage{subfigure}

\let\oldequation\equation
\let\oldendequation\endequation

\renewenvironment{equation}
  {\linenomathNonumbers\oldequation}
  {\oldendequation\endlinenomath}
  
 \let\oldgather\gather
\let\oldendgather\endgather

\renewenvironment{gather}
  {\linenomathNonumbers\oldgather}
  {\oldendgather\endlinenomath}
 
\begin{document}

\title{Generation of high quality sub-two-cycle pulses by self-cleaning of spatiotemporal solitons in air-plasma channels}

\author{Litong Xu}

\author{Tingting Xi}
\email[]{ttxi@ucas.ac.cn}
\affiliation{School of Physical Sciences, University of Chinese Academy of Sciences, Beijing 100049, China}

\begin{abstract}
The temporal sidelobes of few-cycle pulses seriously restrict their applications in ultrafast science. We propose a unique mechanism that enables the generation of sub-two-cycle pulses with high temporal quality based on soliton self-cleaning in air-plasma channels. A robust spatiotemporal soliton could be formed from pulse self-compression by modulating the dispersion of the air-plasma channel via the adjusted plasma density. Due to ionization, the blue-shifted soliton with a larger group velocity captures the leading sidelobes whereas the plasma generated by the soliton defocuses the trailing sidelobes, which are eventually eliminated after a long-distance propagation. The self-cleaning of spatiotemporal soliton leads to no sidelobes in the temporal profile of the sub-two-cycle pulse. The required density of the preformed plasma for arbitrary central wavelength from near-infrared to mid-infrared regime is predicted theoretically and confirmed by (3D+1) simulations.
\end{abstract}

\maketitle
Intense ultrashort pulses down to a few optical cycles have been strongly demanded by ultrafast science \cite{siegrist2019,calegari2014}, especially as important driving sources for attosecond science \cite{sansone2006}.
Few-cycle pulses have been typically generated by pulse compression of supercontinuum based on nonlinear propagation in Kerr media \cite{seo2020,ouille2020,akturk2008,timmers2017,couairon2005,bree2010,skobelev2012}.
The pulse duration has entered single-cycle regime \cite{seo2020} and the peak power has reached Terawatt level \cite{ouille2020}.
However, the temporal quality as another important characteristic has not been improved significantly.
High-peak-power few-cycle pulses, especially sub-two-cycle pulses are usually accompanied by one or several temporal sidelobes, which is very unfavorable for ultrafast applications, such as the generation of isolated attosecond pulses \cite{sansone2006}. 
The sidelobes may be formed due to the uncompensated high-order spectral phase induced by self-steepening \cite{balciunas2015}, ionization \cite{nurhuda2003} or pulse splitting \cite{akturk2008}.
Despite the quality improving approaches including additional third-order dispersion compensation \cite{timmers2017} and pressure gradient \cite{couairon2005}, their parameter sensitivity make it still a challenge to obtain intense few-cycle pulses with high temporal quality.

Recall the spatial mode self-cleaning property of filamentation, where the fundamental mode is focused by Kerr lens to produce high quality filament core, while the higher order modes undergo quick divergence due to weaker focusing and stronger diffraction \cite{liu2007}. If such self-cleaning property can be extended to the temporal domain, we may expect significant improvement of the pulse profile. Simultaneous self-cleaning of both spatial and temporal modes should be related to the formation of long-distance spatiotemporal soliton (STS).
Although STS has been observed in fused silica \cite{durand2013}, it is still unknown how to eliminate the sidelobes of the few-cycle pulse.
If the dispersion of the medium can be adjusted, it may be possible to achieve temporal self-cleaning of STS.
Plasma is a good candidate to provide controllable negative dispersion by changing electron density \cite{kim1990,koprinkov2004}.
To avoid additional defocusing caused by the preformed plasma, we need a plasma channel that is very wide, long, and uniformly distributed.
Recent studies have reported that a long uniform plasma channel could be formed by filamentation of high energy picosecond laser beams in air \cite{tochitsky2019,schmitt-sody2016}.
The spatial scale of plasma channel is much larger than the waist of femtosecond filament, thus it could be a powerful tool to modulate the dispersion relation of air.

In this Letter we propose a new physical scenario that the femtosecond laser pulse self-compresses in a preformed uniform air-plasma channel.
A long-distance STS could be formed under proper negative dispersion of the air-plasma channel, where the required electron density is estimated theoretically and confirmed by simulation results.
The robust propagation of STS leads to pulse self-cleaning, which eliminates sidelobes and significantly improve temporal quality of the pulse.
This scheme can be used to generate sub-two-cycle pulses with different wavelengths, which will greatly favor wavelength-dependent applications in ultrafast science.

Suppose we desire a long-distance STS with a single-cycle duration, two conditions are necessary.
First, the spectrum is super-broadened so that it can support a single-cycle pulse.
Second, the nonlinear length should be close to the dispersion length for a stable temporal soliton, while spatially the beam width may be maintained by the dynamic balance of Kerr focusing and plasma defocusing as a feature of filamentation.
For the first condition, the negative dispersion of air plasma should be large enough to drive the coalescence of the sub-pulses after pulse-splitting, thus synchronizing the spectral components required for a single-cycle pulse.
In our estimation, the two sub-pulses which have a central frequency difference $\Delta \omega$ and time interval $\Delta t$, coalesce after a distance of $\Delta z$ due to negative group-velocity dispersion (GVD):

\begin{equation}
-k_p^{(2)}\Delta\omega\Delta z=\Delta t+k_a^{(2)}\Delta\omega\Delta z.
\end{equation}
Here $k_p^{(2)}$ and $k_a^{(2)}$ are  the GVD coefficients of the plasma and air, respectively. 
The frequency difference of the sub-pulses is supposed to be mainly induced by self-phase modulation, then it is estimated to be $\Delta\omega=2.86B/\tau_0$ \cite{finot2018}, where $\tau_0$ is the full width at half maximum (FWHM) of the initial pulse.
The nonlinear phase shift is taken to be $B=2\pi$, when plasma defocusing becomes  prominent \cite{cheng2016}.
We assume that the two sub-pulses coalesce before the filament diffracts, which means $\Delta z$ is about the diffraction length of filament $\Delta z=k_0w_f^2$.
The radius of filament is $w_f=\lambda_0/\sqrt{8n_2I_p}$, where $n_2$ is the coefficient of nonlinear refractive index \cite{champeaux2005}. 
For simplicity, the clamping intensity is fixed at a typical value of $I_p=60$ TW/cm$^2$.
The time interval $\Delta t$  is assumed to be $\Delta t\approx 0.5\tau_0$.
By applying this setting, we obtain the desired GVD coefficient induced by plasma:
\begin{equation}
\label{k2}
k_p^{(2)}=-\frac{n_2I_p\tau_0^2}{28.2\lambda_0}-k_a^{(2)}.
\end{equation}
From Eq. (\ref{k2}) we can get the corresponding density $\rho_p$ of the preformed plasma channel by $k_p^{(2)}=-1.58\times 10^{-5}\lambda_0^3\rho_p$ [fs$^2$/cm], with $\lambda_0$ and $\rho_p$ in SI units \cite{koprinkov2004}.
We find that the estimated GVD coefficient can also fulfill the second condition, with $L_{n l}=\lambda_0 / 2 \pi n_2 I_p \approx L_{\text {disp }}=\tau_0^2 / 1.665^2\left|k^{(2)}\right|$, where $\tau$ is the FWHM duration of the soliton. 
Then we perform numerical studies on pulse self-compression in air-plasma channels, which verifies our analysis.
 
The propagation of femtosecond laser pulses in air with preformed plasma channel can be described by the nonlinear envelope equation, which governs the pulse envelope $\mathcal{E}(x,y,t,z)$ in the frame moving with the group velocity of central wavelength $(t\rightarrow t-z/v_g )$, coupled with the electron density evolution equation \cite{bree2010}.
The coupled equations are given by:

\begin{gather}
\begin{align}
\frac{\partial\mathcal{E}}{\partial z}=& \frac{i}{2 k_{0}} T^{-1} \nabla_{\perp}^{2} \mathcal{E}+i \frac{\omega_{0}}{c} n_{2} T \int_{-\infty}^{t} \mathcal{R}\left(t-t^{\prime}\right)\left|\mathcal{E}\left(t^{\prime}\right)\right|^{2} d t^{\prime} \mathcal{E} \notag{}\\
&+i \mathcal{D} \mathcal{E}-i \frac{k_{0}}{2\rho_{c0}} T^{-1} \rho \mathcal{E}
-\frac{U_{i} W(|\mathcal{E}|)\left(\rho_{\mathrm{nt}}-\rho\right)}{2|\mathcal{E}|^2} \mathcal{E}
\end{align}
\\\frac{\partial \rho}{\partial t}=W(|\mathcal{E}|)\left(\rho_{\mathrm{nt}}-\rho_0-\rho\right),
\end{gather}
where $z$ is the propagation distance, $k_0=2\pi/\lambda_0$ is the central wavenumber, $T=1+(i/\omega_{0}) \partial_{t}$,  $\mathcal{D}=\sum_{n \geq 2}\left(k^{(n)}/n!\right)\left(i \partial_{t}\right)^{n}$, $k^{(n)}=\partial^{n} k /\left.\partial \omega^{n}\right|_{\omega_{0}}$.
 The dispersion relation of air-plasma channel is $k(\omega)=k_a(\omega)-\rho_0\omega/2c\rho_c$.
Here $k_a(\omega)$ is the dispersion relation of air \cite{zhang2008,Mathar2007} and $\rho_0$ is the density of preformed plasma.
 The critical plasma density $\rho_c(\lambda)=1.11\times 10^9/\lambda^2\ \mathrm{cm}^{-3}$ is wavelength dependent, accounting for the negative dispersion, and $\rho_{c0}=\rho_c(\lambda_0)$ .
The ionization rate $W(|\mathcal{E}|)$ is calculated from the Perelomov-Popov-Terent’ev model \cite{ppt}.
The ionization potential $U_i=12.1$ eV, and the neutral oxygen density $\rho_{nt}=5.4\times 10^{18}$ cm$^{-3}$.
$\mathcal{R}\left(t-t^{\prime}\right)$ is the nonlinear Kerr response including a Raman-delayed contribution \cite{berge2008b}, and the nonlinear refractive index $n_2=1.3\times 10^{-19}\ $cm$^2$/W \cite{berge2008}.
The input laser pulse has a Gaussian spatiotemporal distribution with a temporal duration (FWHM) $\tau_0$, and a beam waist $2w_0=2$ mm, focused by a lens with $f=1$ m.
The input peak power is $P_{\mathrm{in}}=4P_{\mathrm{cr}}$, where the critical power for self-focusing is $P_{\mathrm{cr}} = 3.77 \lambda_0^2 / 8 \pi n_0 n_2$ \cite{tochitsky2019}.

 \begin{figure}[tb]
 \includegraphics[width=0.5\textwidth]{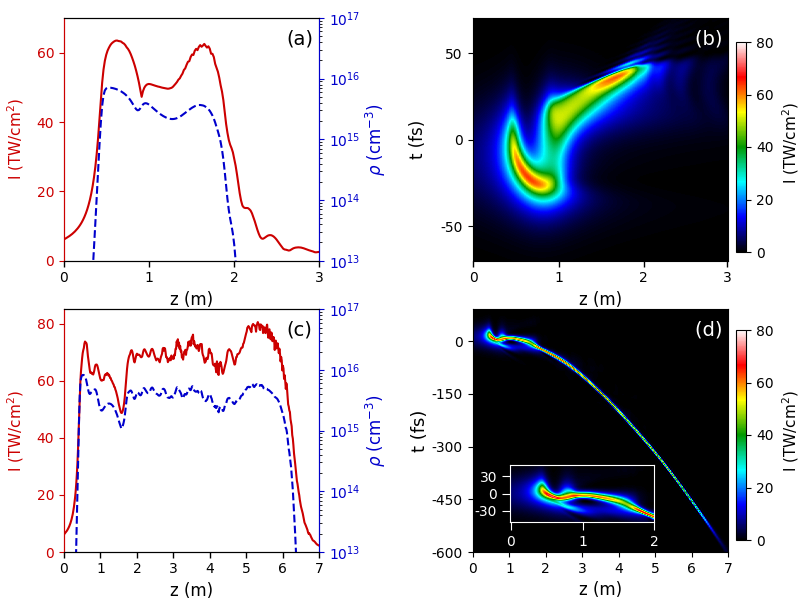}
\caption{\label{fig1} (a)(c) Peak intensity (solid lines) of the femtosecond pulse and the generated electron density (dashed lines) as a function of propagation distance, when the 1.5-$\mu$m pulse propagates in (a) air and (c) air-plasma channel with $\rho_0=8.8\times 10^{16}\ \mathrm{cm}^{-3}$. (b)(d) On-axis pulse evolution when propagating in (b) air and (d) air-plasma channel.}
 \end{figure} 
 
 We start with the case when $\lambda_0=1.5\ \mu$m and $\tau_0=50$ fs.
Figure \ref{fig1}(a) shows the peak intensity evolution of the 1.5-$\mu$m pulse propagating in air and air-plasma channel with the required electron density $\rho_p=8.8\times10^{16}\ \mathrm{cm}^{-3}$ predicted by  Eq. (\ref{k2}).
For the propagation in air, as the pulse self-focuses, the intensity increases rapidly and then is clamped at 60 TW/cm$^2$.
Accordingly, the electron density generated by the pulse is increased to $7\times 10^{15}$ cm$^{-3}$.
As a result, the pulse splits at $z=0.8$ m due to plasma defocusing and subsequent refocusing of the pulse trailing edge, as shown in Fig. \ref{fig1}(b).
With the diffraction of the sub-pulses, the filamentation is terminated at about $z=2$ m.
Compared to the propagation in air, the filament is prolonged more than threefold when there is a preformed plasma channel [Fig. \ref{fig1}(c)].
In this case, the filamentation process is divided into two stages at $z=1.7$ m.
In the first stage, the intensity evolution is similar to that without preformed plasma channel, with a clamping intensity $\sim 60$ TW/cm$^2$.
In the beginning of the second stage, the intensity first decreases due to pulse-splitting and then increases again due to the coalescence of sub-pulses. 
The coalescence of sub-pulses can enrich the spectrum, making it able to support a shorter pulse.
Then the intensity is clamped at $\sim 70$ TW/cm$^2$ from $z=2$ m to 6 m.
At this stage, a stable temporal profile is observed [Fig. \ref{fig1}(d)].
Combined with the spatial characteristics of filamentation, the STS is generated as we expect.

  \begin{figure}[tb]
 \includegraphics[width=0.5\textwidth]{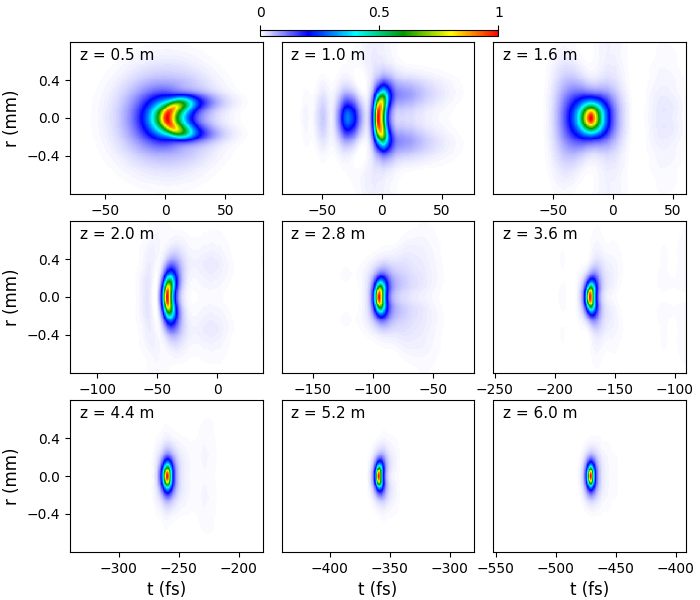}%
 \caption{\label{fig2} Spatiotemporal intensity distribution at several propagation distances for
 $\lambda_0=1.5\ \mu$m, $\tau_0=50$ fs, $\rho_0=8.8\times 10^{16}\ \mathrm{cm}^{-3}$.}
 \end{figure} 

\begin{figure}[tb]
 \includegraphics[width=0.5\textwidth]{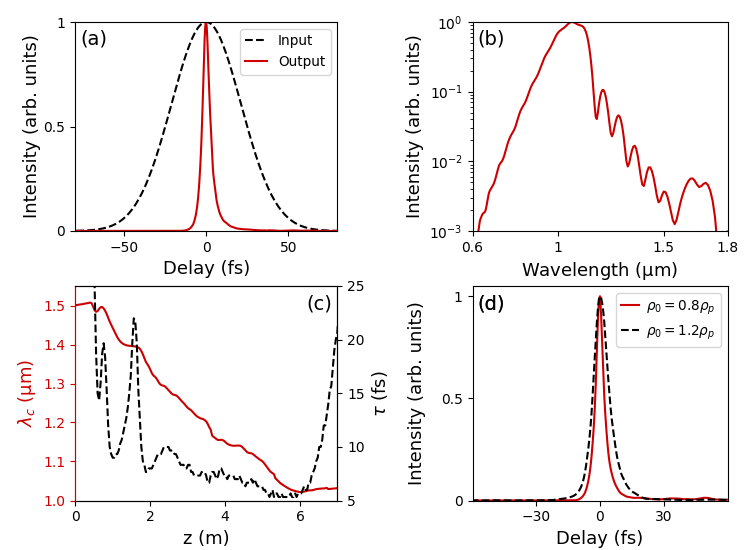}%
 \caption{\label{fig3} (a) Output pulse profile and (b) spectrum at $z = 6$ m when $\rho_0=\rho_p=8.8\times 10^{16}\ \mathrm{cm}^{-3}$, the intensity is integrated over a circular region with 200 $\mu$m radius. (c) The evolution of carrier wavelength $\lambda_c$ (solid line) and the pulse duration $\tau$ (dashed line). (d) Output pulses when employing air-plasma channels with $\rho_0 = 0.8\rho_p$ (solid line) and $\rho_0 = 1.2\rho_p$ (dashed line).}
 \end{figure}

To observe detailly the propagation of STS and confirm its self-cleaning, in Fig. \ref{fig2} we plot the spatiotemporal intensity distribution of the pulse at several propagation distances.
Such a long-distance STS is crucial to pulse self-cleaning.
At $z=0.5$ m, the pulse trailing edge is defocused by the plasma generated at the leading edge, leading to a fishbone structure in the intensity distribution.
Then we see pulse-splitting at $z=1$ m.
After $z=1$ m, the rear sub-pulse with higher frequency starts to catch up with the front pulse.
Consequently, the coalescence of the two sub-pulses occurs at $z=1.6$ m. 
The isolated pulse with duration of 22 fs continues self-compression, which is compressed to 8 fs at $z = 2$ m.
At the initial stage of soliton regime, there are sidelobes both before and after the main pulse.
Both ionization and the long-distance propagation play important roles in eliminating these sidelobes.
As shown in Fig. \ref{fig3}(c), the carrier wavelength continuously shifts towards blue side due to ionization.
As a result, the main pulse with a larger group velocity captures the front sidelobes.
Meanwhile, the plasma generated by the STS has a density as high as $5\times 10^{15}\ \mathrm{cm}^{-3}$ [Fig. \ref{fig1}(a)], leading to strong plasma defocusing of trailing sidelobes.
The power that the latter sidelobes possess is insufficient to overcome plasma defocusing, thus the latter sidelobes are gradually dissipated to off-axis region.
Therefore, the sidelobes are eliminated and the temporal profile of the few-cycle pulse is self-cleaned after a long-distance propagation.

As a further confirmation of this fact, the output pulse profile at $z = 6$ m, integrated over a circular region with 200 $\mu$m radius, is shown in Fig. \ref{fig3}(a).
The pulse duration (FWHM) is 5.3 fs, indicating a compression factor close to 10.
More importantly, the sub-two-cycle pulse has a clean pulse shape without accompanying sidelobes.
Such optimization in the temporal quality will be particularly beneficial to the generation of isolated attosecond pulses.
Note that the corresponding spectrum is quite different from commonly seen supercontinuum from filamentation, as shown in Fig. \ref{fig3}(b). The spectrum spans more than an octave, with the carrier wavelength shifted to $\sim 1.1\ \mu$m due to ionization. To detail the evolution of the soliton, in Fig. \ref{fig3}(c) we plot the evolution curves of the carrier wavelength $\lambda_c=\int_0^{\infty}I(\lambda)\lambda d\lambda/\int_0^{\infty}I(\lambda) d\lambda$ and pulse duration. 
Starting from about $z = 1$ m, the carrier wavelength decreases with almost a fixed rate, which may greatly facilitate frequency tuning of the desired pulse. 
At the soliton stage (after $z=2$ m),  the pulse duration continuously decreases.
This can be understood by the self-organized behavior of the soliton. 
In order to maintain the equilibrium between Kerr effect and dispersion, the soliton  will continue self-compression to compensate for the energy dissipation caused by ionization.
When the soliton cannot be sustained (after $z=6$ m), the laser intensity drops, and the pulse is stretched rapidly.

It should be noted that due to the self-organized behavior of the STS,  slightly changing the dispersion condition will not prevent the formation of STS.
Figure \ref{fig3}(d) shows the output pulse profiles when there is a 20\% deviation of the plasma density.
When $\rho_0 = 0.8\rho_p$, the pulse is compressed to 4.3 fs, but there is a low-intensity sidelobe after the main pulse, which results from the shortened soliton propagation length.
When $\rho_0 = 1.2\rho_p$, there is no sidelobe, but the pulse duration is larger (8.2 fs).
This is because for stronger dispersion, only a relatively large pulse duration can maintain the stability of the soliton, which is also consistent with the temporal soliton solution in Ref. \cite{tzoar1981}.

 % If in two-column mode, this environment will change to single-column
% format so that long equations can be displayed. Use
%\begin{widetext}
%\end{widetext}

  \begin{figure}[tb]
 \includegraphics[width=0.5\textwidth]{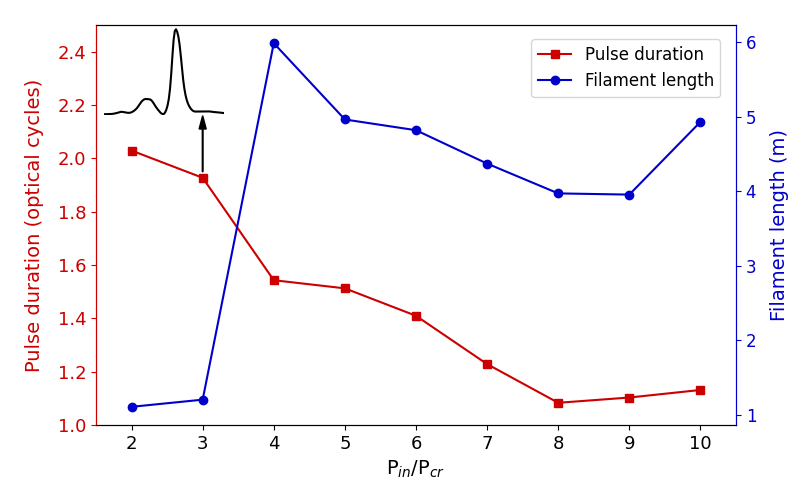}%
 \caption{\label{fig4}The output pulse duration and filament length as a function of input power. The pulse duration is in the unit of optical cycles, $\rho_0$ is fixed at $8.8\times 10^{16}\ \mathrm{cm}^{-3}$. The inset shows the output pulse profile when $P_{\mathrm{in}}=3P_{\mathrm{cr}}$.}
\end{figure}

We also investigate the pulse self-compression for different input laser power. 
 Figure \ref{fig4} shows the output pulse durations when $P_{\mathrm{in}}$ changes from 2$P_{\mathrm{cr}}$ to 10$P_{\mathrm{cr}}$.
 The filament length, defined as propagation distance that the maximum intensity is larger than 30 TW/cm$^2$, is given in the same plot.
 When $P_{\mathrm{in}}<4P_{\mathrm{cr}}$, the spectrum is not wide enough after pulse splitting, and the sub-pulses cannot fully coalesce before intensity drops. No STS is formed and the filament length is only $\sim$ 1 m.
 Therefore, the output pulse duration is larger, and the sidelobes are observed.
 For $P_{\mathrm{in}}=4P_{\mathrm{cr}}$, the filament length is significantly extended. 
 In this case, long-distance STS eliminates temporal sidelobes of the pulse as discussed before.
 As $P_{\mathrm{in}}$ increases from $4P_{\mathrm{cr}}$ to $8P_{\mathrm{cr}}$ , the filament length keeps in $4-5$ m, and the output pulse duration slightly decreases, close to single-cycle regime.
 This is because for higher inout power, plasma defocusing is stronger before pulse-splitting.
 Then at the refocusing stage, the spatiotemporal focusing effect leads to higher clamping intensity and thus broader spectrum.
 Further increasing $P_{\mathrm{in}}$ cannot increase the clamping intensity anymore.
 Therefore, the pulse duration is limited to single-cycle regime.
 Moreover, simply increasing the input power cannot significantly increase the output peak power ($\sim 50$ GW) due to intensity clamping and soliton size limitation.

  \begin{figure}[tb]
 \includegraphics[width=0.5\textwidth]{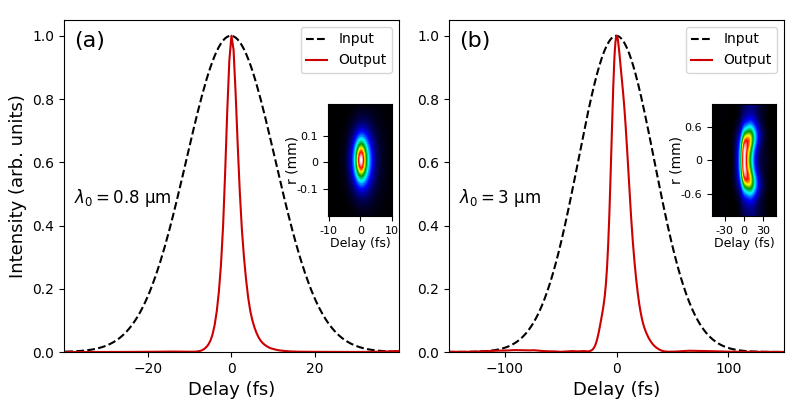}%
 \caption{\label{fig5}Input (dashed lines) and output (solid lines) pulse shapes when applying the plasma compression scheme to (a) Near-infrared and (b) Mid-infrared laser pulses. Simulation parameters: (a) $\lambda_0=0.8\ \mu$m, $\rho_0=2.9\times 10^{17}\ \mathrm{cm}^{-3}$, $\tau_0=25$ fs; (b)  $\lambda_0=3\ \mu$m, $\rho_0=1.4\times 10^{16}\ \mathrm{cm}^{-3}$, $\tau_0=80$ fs.
 The input peak power is $P_{\mathrm{in}}=4P_{\mathrm{cr}}$.
 The output pulse durations are integrated over circular regions with radius (a) $r_0=100\ \mu$m and (b) $r_0=600\ \mu$m.
The insets show the spatiotemporal intensity distribution.}
 \end{figure}

 Since the STS results from the interplay of nonlinear effects and GVD, our scheme should be applicable to pulses with different central wavelengths by properly modulating the dispersion relation of air.
 As a check on this, we tested our scheme for 0.8-$\mu$m and 3-$\mu$m pulses.
The input pulse durations are 25 fs for $\lambda_0=0.8\ \mu$m and 80fs for $\lambda_0=3\ \mu$m, and the input power is kept at $P_{\mathrm{in}}=4P_{\mathrm{cr}}$.
Using the predicted values of plasma density from Eq. (\ref{k2}), $\rho_p=2.9\times 10^{17}$ and $1.4\times 10^{16}\ \mathrm{cm}^{-3}$, respectively, we successfully obtain sub-two-cycle pulses with high temporal quality, as shown in Fig. \ref{fig5}.
The output pulse profiles are integrated over circular regions with radius $r_0=100\ \mu$m for  $\lambda_0=0.8\ \mu$m and $r_0=600\ \mu$m for  $\lambda_0=3\ \mu$m, giving pulse durations of 3.7 fs  and 17.3 fs, both shorter than two optical cycles.
Similar STS is also observed in the spatiotemporal intensity distribution [insets of Fig. \ref{fig5}], which further confirms that our scheme can be applied to generate few-cycle pulses with desired wavelength.
The versatility of our scheme should greatly facilitate wavelength-dependent applications of the few-cycle pulses in ultrafast science.
 
In summary, we have proposed a new physical scenario that the femtosecond laser pulse self-compresses in a preformed uniform air-plasma channel, and demonstrated its significant potential for  generating sub-two-cycle pulses with high temporal quality.
Plasma-induced negative dispersion enables the generation of a long-distance spatiotemporal soliton, which leads to continuous pulse self-cleaning. 
This scheme can be used to compress femtosecond laser pulses with different central wavelengths to sub-two-cycle pulses with high spatiotemporal quality by changing the density of preformed plasma channel.

\nolinenumbers
\begin{acknowledgments}
The authors would like to thank Prof. Jie Zhang of Shanghai Jiao Tong University for fruitful suggestions. This work was supported by the National Natural Science Foundation of China (Grants No. 11874056, No. 12074228, No. 11774038).
\end{acknowledgments}

\end{document}